\documentclass[a4paper, 11pt]{article} 
\usepackage{graphicx}      
\usepackage{float}	
\usepackage{tikz}
\usepackage{amsmath}
\usepackage{amsfonts}
\usepackage{amssymb}
\usepackage{subfig}
\usepackage{dblfloatfix}
\usepackage{scalefnt}
\usepackage[linesnumbered, ruled]{algorithm2e}
\usepackage[top=2cm, bottom=2cm, left=3cm, right=3cm]{geometry}
\newtheorem{remark}{Remark}

\title{\normalsize Appendix to \cite{kergus2019}:\\ \Large \bf
Data-driven reference model selection \\ and application to L-DDC design
}

\author{Pauline Kergus$^{1\star}$, Martine Olivi$^{2}$, Charles Poussot-Vassal$^{1}$ and Fabrice Demourant$^{1}$
\thanks{Thanks to the whole FACTAS team at INRIA and particularly to Adam Cooman for his precious advice on data-driven stability analysis.}
\thanks{$^{1}$ONERA, DTIS, Information Processing and Systems, Toulouse, France}%
\thanks{$^{2}$FACTAS team, INRIA Sophia Antipolis, France}%
\thanks{$^\star$Corresponding author: \tt \small{pauline.kergus@onera.fr}}
}

\begin{document}
\maketitle
\thispagestyle{empty}
\pagestyle{empty}

\tikzstyle{sum}=[draw,circle,text width = 0.3cm]
\tikzstyle{block}=[rectangle, draw=black, text centered]

\section{Introduction}

For many applications, a mathematical description of the system, derived from physical laws, is not available. In this case, the controller has to be designed on the basis of experimental measurements. The first solution consists in identifying a model of the plant and then using any kind of model-based technique to obtain a control law (indirect methods). It is indicated for problems where a reliable model with bounded modeling errors is available. On the other side, the data-driven strategy directly computes the controller from the experimental data. Such techniques are also called direct methods and may be appealing in the cases such a control-oriented model is too time-consuming, too complex or too costly to obtain. The two strategies, model-based and data-driven, are complementary in sense that they do not address the same categories of problem.
\par \leavevmode \par
\begin{figure}[H]
\centering
	\scalebox{1.2}{
	\begin{tikzpicture}
	\draw
	
	node at (0,0)[]{}
	node [name=ref] {} 
	
	node at (-0.5,0)[](r1){}
	node at (2,0.9)[block, fill=blue!10](M){M}
	node at (2,1)[](y){}
	
	node at (1,0)[sum](comp){}
	node at (2,0)[block, fill=red!10](cor){K}
	node at (3,0)[block](plant){P}
	node at (4,0)[](y1){}
	
	node at (4.5,0)[sum](comp2){};
	
	\draw (comp.135) -- (comp.315);
    \draw (comp.225) -- (comp.45);
    \node (plus) [below] at (comp.180) {$+$};
    \node (minus) [below] at (comp.350) {$-$};
    
    \draw (comp2.135) -- (comp2.315);
    \draw (comp2.225) -- (comp2.45);
    \node (plus) [below] at (comp2.180) {$-$};
    \node (minus) [above] at (comp2.50) {$+$};
	
	\draw[->](r1) -- node[]{}(comp);
	\draw[->]node[above]{}(-0.1,0) |- (M) -| node[below]{}(comp2);
	\draw[->](comp) -- node[above]{}(cor) -- node[above]{} (plant) -- node[above]{}(comp2);
	\draw[->](3.5,0) |- (1,-1) -- node[right]{}(comp);
	\draw[-, dash pattern=on 2pt off 1pt, line width=0.5mm, blue!50] (0.2,-1.2) -- (0.2,0.5) -- (4,0.5)--(4,-1.2) -- (0.2,-1.2);
	\draw[->](comp2) -- node[above]{$\varepsilon$}(5.5,0);

	\end{tikzpicture}}
	
	\caption{Problem formulation: $M$ is the desired closed-loop, $P$ is the plant and $K$ the controller to be designed.}
	\label{problem_formulation}
	\vspace{-0.5cm}
\end{figure}
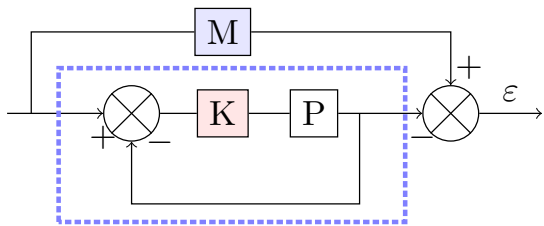

  
Numerous direct methods have been proposed to try to achieve the best possible performance without using any plant model. Among them, some methods, like Iterative Feedback Tuning (IFT, \cite{hjalmarsson1998iterative}), Correlation-based-Tuning (CbT, \cite{karimi2002convergence}), Virtual Reference Feedback Tuning (VRFT, \cite{campi2002virtual}) or Loewner Data-Driven Control (LDDC, \cite{kergus2017}), can be designated as model-reference techniques. The principle of the model-reference problem is recalled on Figure \ref{problem_formulation}. These approaches only require data from the plant $P$ and the desired closed-loop behaviour, given as a reference model transfer function $M \in \mathcal{RH}_\infty$. The objective is to design a controller that  minimizes the error $\epsilon$ between the resulting closed-loop and the desired one $M$.

\par \leavevmode \par  
The choice of a reference model in data-driven control techniques is a critical step, see \cite{kergus2019}, \cite{bazanella2011data}, \cite{piga2018direct}, \cite{selvi2018towards}. Indeed, it should represent the desired closed-loop performances and be achievable by the plant at the same time. In \cite{kergus2019}, a method to build such a reference model, both reproducible by the system and having a desired behaviour, is proposed. It is applicable to Linear Time-Invariant (LTI) monovariable systems. The present paper aims at providing more applications of this method. The LDDC (Loewner Data Driven Control) algorithm is used to illustrate the impact of the choice of the reference model on the control design process.
\par \leavevmode \par 
This paper is an appendix of \cite{kergus2019}. Its objective is to provide additional examples. This article is organized in five sections. Section \ref{algo} recalls the method proposed in \cite{kergus2019}. Two control applications are given in Sections \ref{appli_cry} and \ref{appli_EDF}. The first considered plant is a continuous crystallizer. The second one is an open channel for hydroelectricty generation. Finally, conclusions and outlooks are proposed in Section \ref{conclusion}.

\section{Preliminary: Data-driven selection of a reference model}
\label{algo}
\subsection{Definition of an achievable closed-loop}
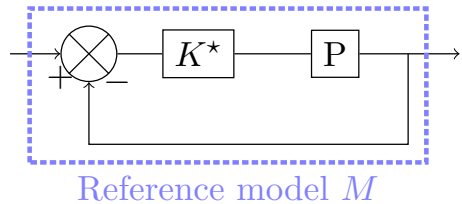
\begin{figure}[H]
\centering
\scalebox{1.2}{
\begin{tikzpicture}
	\draw
	
	node at (0,0)[]{}
	node [name=ref] {} 
	node at (1,0)[sum](comp){}
	node at (2.2,0)[block](cor){$K^\star$}
	node at (3.7,0)[block](plant){P}
	node at (5.2,0)[](y1){};

	\draw (comp.135) -- (comp.315);
    \draw (comp.225) -- (comp.45);
    \node (plus) [below] at (comp.180) {$+$};
    \node (minus) [below] at (comp.350) {$-$};
	
	\draw[->](ref) -- node[below]{}(comp);
	\draw[->](comp) -- node[above]{}(cor) -- node[above]{}(plant) --node[above]{}(y1);
	\draw[->](4.5,0) -- (4.5,-1) -| node[below]{}(comp);
	\draw[-, dash pattern=on 2pt off 1pt, line width=0.5mm, blue!50] (0.35,-1.2) -- (0.35,0.5) -- (4.7,0.5)--(4.7,-1.2)--node[below]{Reference model $\color{blue!50}{M}$}(0.35,-1.2);
	\end{tikzpicture}}
    \caption{The ideal case: feedback interconnection obtained with the ideal controller.}
    \label{ideal_case}
\end{figure}

By achievable behaviour, we mean that the reference model corresponds to an internally stable closed-loop interconnection, see Figure \ref{ideal_case}. In other words, the ideal controller $K^\star$ should stabilize the plant internally. It is well known that the interconnection visible on Figure \ref{ideal_case} is internally stable if and only if the sensitivity function $1-M$ is stable and if there are no compensation of instabilities in the open-loop, between the ideal controller $K^\star$ and the plant $P$.

\par \leavevmode \par
Therefore, as explained in \cite{kergus2019}, the reference model $M$ is achievable by the plant if and only if $M$ is stable and satisfies:
\begin{equation}
    \left\{\begin{array}{l}
        \forall i=1\dots n_z, \ M(z_i) = 0 \\
        \forall j=1\dots n_p, \  M(p_j) = 1 
    \end{array}\right. ,
    \label{defM}
\end{equation}
where $\left\{z_i\right\}_{i=1}^{n_z}$ and $\left\{p_j\right\}_{j=1}^{n_p}$ are respectively the unstable zeros and poles of the plant, which are assumed to be distinct. These interpolatory conditions are equivalent to having no compensation of instabilities between the ideal controller $K^\star$ and the plant $P$. When the plant has multiple RHP poles or zeros, derivative constraints must be added to \eqref{defM}, see \cite{kergus2019} for further information.

\subsection{Data-driven selection of an achievable reference model}
The method proposed in \cite{kergus2019} is briefly summed up in Algorithm \ref{algo_choixM}. Details about the different steps are given in the following paragraphs. 
\begin{algorithm}
\SetAlgoLined
\SetKwInOut{Solution}{Solution}
\KwData{\begin{itemize}
    \item Samples of the frequency response of the plant $\{\omega_{i},P(\imath\omega_i)\}, \ i=1 \ldots N$.
    \item Stable reference model $M$ giving the desired closed-loop performances.
\end{itemize}}
\Solution{
\begin{enumerate}
    \item Project the frequency-response of the plant $P$ on the Hardy spaces $\mathcal{H}_2$ and $\overline{\mathcal{H}}_2$ and determine whether the plant is stable or not, see Section \ref{stab_analysis}. If the plant is unstable, estimate its RHP poles.
    
    \item As in step 1, perform the projection of the frequency response $\{\omega_{i},P(\imath\omega_i)^{-1}\}, \ i=1 \ldots N$ of $P^{-1}$ to determine if the plant is minimum phase or not. If the plant is non-minimum phase, estimate its RHP zeros.
    
    \item According to the nature of the plant, select an achievable reference model:
    \begin{enumerate}
        \item \textbf{If the plant is stable and minimum phase}: any stable and minimum-phase $M$ specified by the user is achievable by the plant.
        \item \textbf{If the plant is stable and non-minimum phase}: $M_f=MB_z$ is achievable by the plant with 
        $$B_z(s)= \prod_{i=1}^{n_z}\frac{s-z_i}{s+z_i} $$
        where $\left\{z_i\right\}_{i=1}^{n_z}$ are the RHP zeros of the plant estimated at Step 2.
        \item \textbf{If the plant is unstable and minimum phase}: $M_f=1-(1-M)B_p$ is achievable by the plant with
        $$B_p(s) = \prod_{j=1}^{n_p}\frac{s-p_j}{s+p_j}$$
        where $\left\{p_j\right\}_{j=1}^{n_p}$ are the RHP poles of the plant estimated at Step 1.
        \item \textbf{If the plant is unstable and non-minimum phase}: $M_f=MB_zF$ is achievable by the plant, where the filter $F$ is defined as follows:
        $$F(s)=\frac{\sum_{k=1}^{n_p}\gamma_k l_k(s)}{\prod_{j=1}^{n_p} (s+p_j)},$$
        with $\gamma_k=\frac{\prod_{j=1}^{n_p} (p_k+p_j)}{M(p_k)B_z(p_k)}$ and $l_k(s)=\prod_{j=1, j\ne k}^{n_p}\frac{s-p_j}{p_k-p_j}$, for $k=1\dots n_p$.
    \end{enumerate}
    
\end{enumerate}
}
\caption{Data-driven selection of an achievable reference model}
\label{algo_choixM}
\end{algorithm}

\subsubsection{Data-driven stability analysis}
\label{stab_analysis}
In order to define the interpolatory conditions of \eqref{defM} for a given plant $P$, it is necessary to determine its RHP poles and zeros. This can be done in a data-driven way using the methods presented in \cite{cooman2018estimating} and \cite{cooman2018model}.
\par \leavevmode \par 
The first step consists in a projection of the FRF measurements on the space $\mathcal{H}_2$ and $\overline{\mathcal{H}}_2$:
\begin{equation}
    P(\imath\omega)=P^{s}(\imath \omega)+P^{as}(\imath \omega)
    \label{proj_data}
\end{equation}
where $P^{s} \in \mathcal{H}_2$ is its stable projection while $P^{as} \in \overline{\mathcal{H}}_2$ is its anti-stable part. On the basis of \eqref{proj_data}, it is possible to determine if the plant is stable or not. For further information on the projection method, see \cite{cooman2018model}.
\par \leavevmode \par
If the plant is unstable, it is then possible to estimate its RHP poles using its anti-stable projection $P^{as}$ as explained in \cite{cooman2018estimating}. These techniques are implemented in the PISA Toolbox \cite{PISA}.

\subsubsection{Selection of an achievable reference model}
As said earlier, an achievable reference model is stable and satisfies the interpolatory conditions given in \eqref{defM}, which are now well defined thanks to the estimation of the plant's instabilities at the previous step.
\par \leavevmode \par 
The selection of a reference model depends on the nature of the plant. The proposed choice is detailed and justified in \cite{kergus2019}.

\section{Application to a continuous crystallizer}
\label{appli_cry}
The  first considered application is the control of a continuous cooling crystallizer. This process is widely used in the chemical industry. It is a separation process which goal is to produce high-purity solids from liquids. The system is SISO: its input is the solute feed concentration $c_f(t)$ and its output is the solute concentration in the crystallizer $c(t)$. The state of the system is $x(t)=\left[n(L,t) \ c(t)\right]^T$, where $n(L,t)$ denotes the crystal size distribution. Physically, this system is described by population and mass balance equations. A complete mathematical model of this system is derived in \cite{rachah2016mathematical}.
\par \leavevmode \par 
The objective is to stabilize the plant around a desired steady-state $c(t)=c_{ss}=4.09 mol/L$, which is just above the saturation concentration $c_s=4.038mol/L$, required for the crystals to be produced. For this steady state, as said in \cite{vollmer2001h}, the system is unstable and presents sustained oscillations which may degrade the crystals quality. Feedback control is therefore needed. This control problem has been treated in \cite{vollmer2001h} through infinite-dimensional $\mathcal{H}_\infty$ synthesis, which is model-based, and in \cite{apkarian2017structured} thanks to a data-driven $\mathcal{H}_\infty$ synthesis.
\par \leavevmode \par
\begin{figure}
    \centering
    \includegraphics[width=0.75\textwidth]{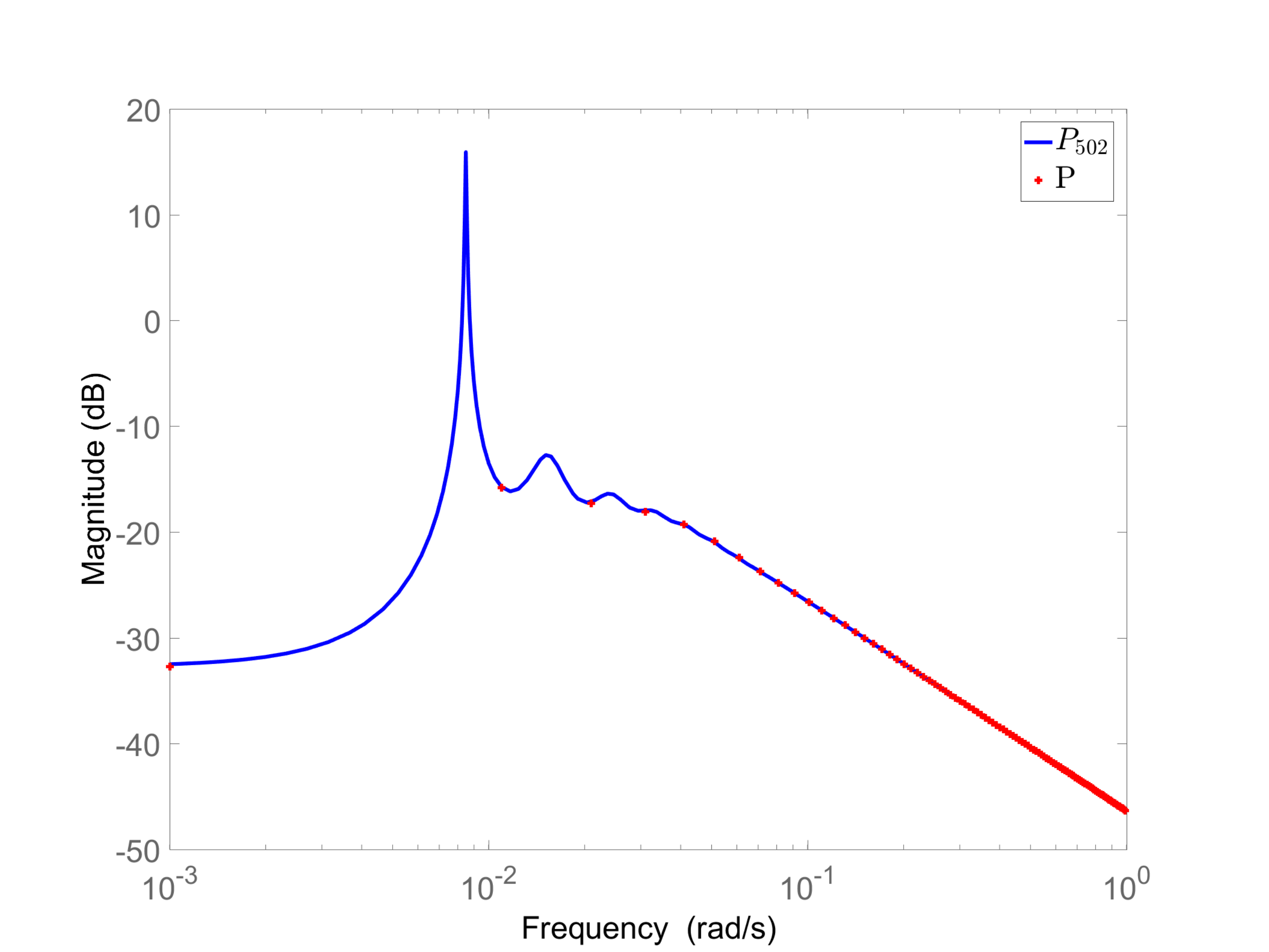}
    \caption{Evaluation of the frequency-response of the linearized plant $P$ on a fine frequency grid ($10^5$ linspaced frequencies between 0.001 and 1000 rad.s$^{-1}$) and frequency response of the finite difference rational model $P_{502}$.}
    \label{cry_frf}
\end{figure}

\begin{figure}
    \centering
    \includegraphics[width=0.75\textwidth]{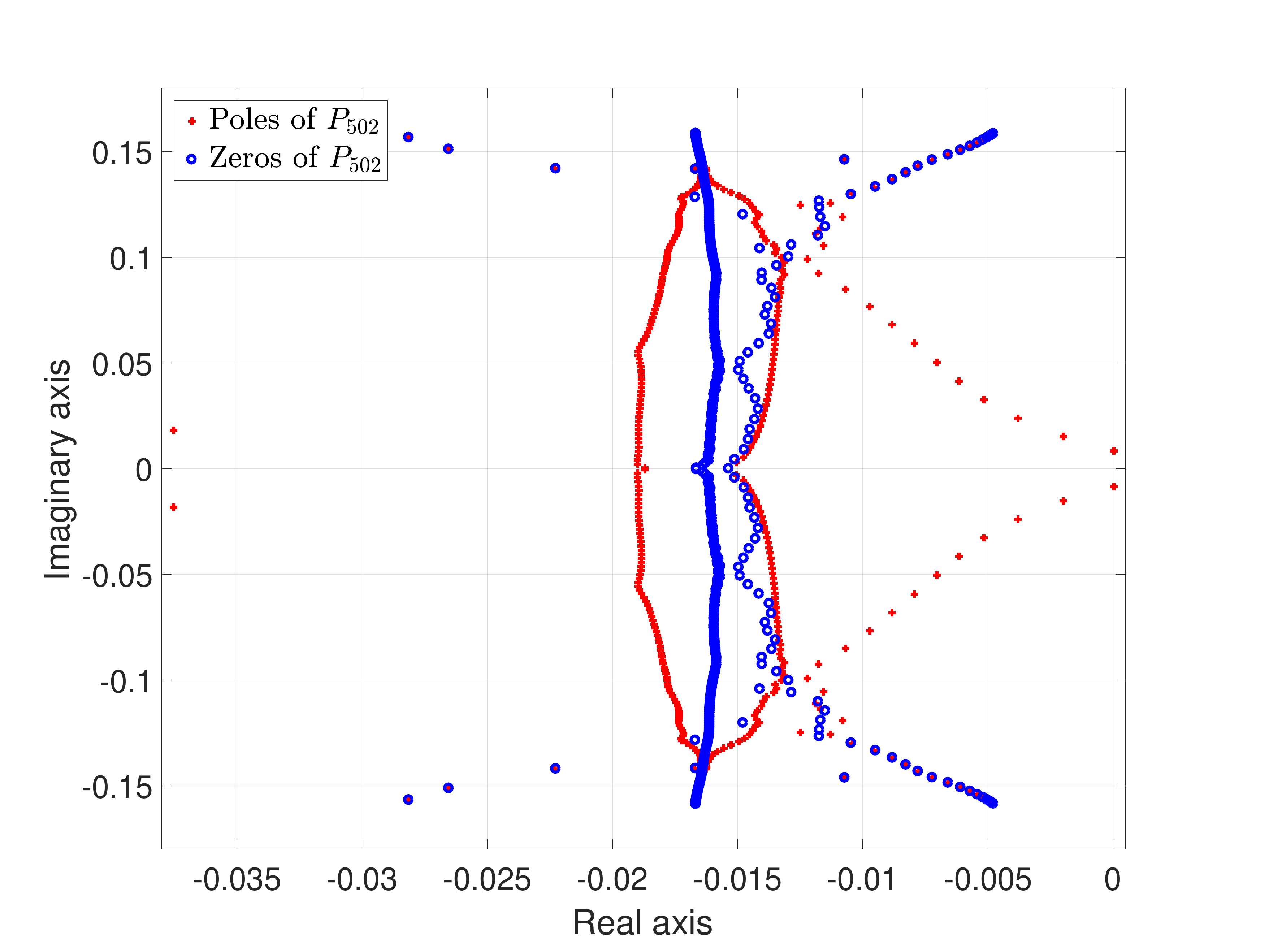}
    \caption{Poles and zeros of the rational model $P_{502}$}
    \label{pzmap_cry}
\end{figure}

 
When linearizing the system's partial differential equations around the desired steady state, the crystallizer is characterized by an irrational transfer with an infinite number of poles, see \cite{vollmer2001h}, \cite{rachah2016mathematical} or \cite{apkarian2017structured} for its expression. It is possible to evaluate numerically the frequency response of the linearized plant $P$ on a discrete frequency grid, see Figure \ref{cry_frf}. As in \cite{apkarian2017structured}, a rational model $P_{502}$ of order 502 is obtained through a finite-difference method (see Figure \ref{cry_frf}). The poles and zeros of $P_{502}$ are given on Figure \ref{pzmap_cry}: the rational model is minimum phase and two unstable poles of value $3.83\times10^{-5}\pm 0.848\times 10^{-2}\imath$ are visible. 
\par \leavevmode \par 
In order to use the proposed method to select an achievable reference model, the considered frequency grid is much smaller: $N=500$ frequencies are considered, logspaced between $10^{-3}$ and 1 rad.s$^{-1}$. The corresponding samples of the frequency response of the plant are estimated directly through the irrational transfer.
\par \leavevmode \par
\begin{figure}[H]
    \centering
    \includegraphics[width=0.75\textwidth]{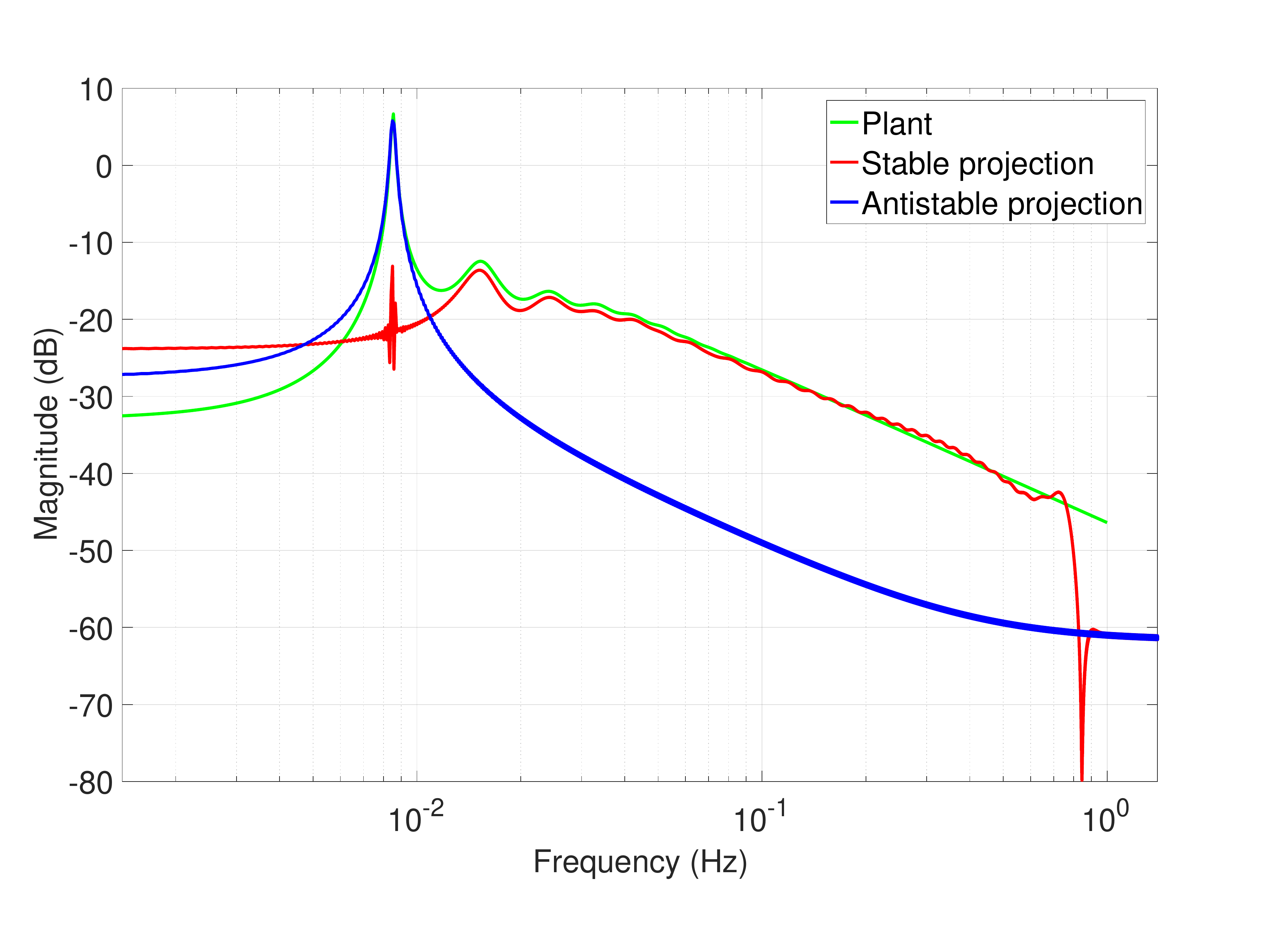}
    \caption{Projection of the frequency-response samples from the plant $P$ on the Hardy spaces $\mathcal{H}_2$ and $\overline{\mathcal{H}}_2$: the system is unstable.}
    \label{proj_cry}
\end{figure}

 
The first step of Algorithm \ref{algo_choixM} consists in the projection of the FRF measurements on the spaces $\mathcal{H}_2$ and $\overline{\mathcal{H}}_2$. The projection is given on Figure \ref{proj_cry}: the antistable projection fits the resonance while the stable part fits the rest of the frequency-response of the plant. Therefore the plant is unstable, as expected considering $P_{502}$ but also the previous studies of this system given in \cite{vollmer2001h}, \cite{rachah2016mathematical} and \cite{apkarian2017structured}.
\par \leavevmode \par
\begin{figure}[H]
    \centering
    \includegraphics[width=0.75\textwidth]{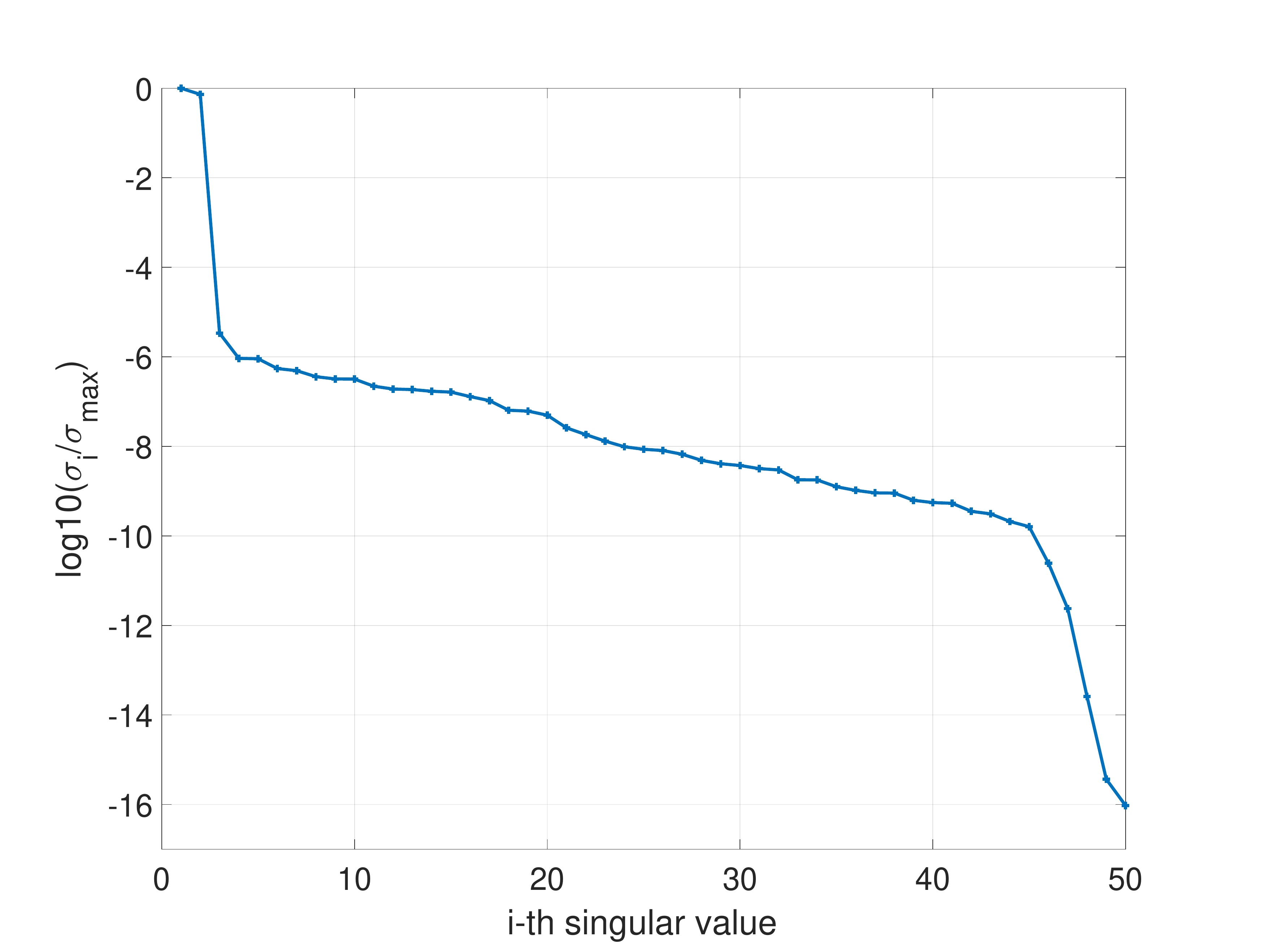}
    \caption{Singular Value Decomposition of the Hankel matrix corresponding to the antistable projection of the plant's data: the system exhibits two RHP poles.}
    \label{svd_Hankel}
\end{figure}

 
As detailed in \cite{cooman2018estimating}, the Hankel matrix of the antistable projection of the plant's data is computed and a singular value decomposition is performed. The rank of this matrix gives us the order of the antistable projection, and consequently the number of RHP poles of the system $P$. The decomposition is visible on Figure \ref{svd_Hankel}: according to the drop after the second singular value, the system exhibits two unstable poles.
These two RHP poles are then estimated. Their value is given in Table \ref{RHPpoles}. The obtained values are coherent with the ones found in \cite{vollmer2001h} and with the RHP poles of the rational model $P_{502}$.
\begin{table}[H]
    \centering
    \begin{tabular}{|c|c|}
        \hline
        RHP poles of $P_{502}$ & $3.83\times10^{-5}\pm 0.848\times 10^{-2}\imath$ \\
        \hline
        Estimated RHP poles & $1.07\times10^{-4}\pm 0.852\times 10^{-2}\imath$ \\
        \hline
        Estimated RHP poles in \cite{vollmer2001h} & $0.99\times10^{-4}\pm 0.89\times 10^{-2}\imath$\\
        \hline
    \end{tabular}
    \caption{Estimation of the RHP poles of the plant.}
    \label{RHPpoles}
    \vspace{-0.15cm}
\end{table}

\begin{remark}
In \cite{vollmer2001h}, the RHP poles are estimated through a direct search method. This estimation is then used to factorize the plant's expression to solve the mixed-sensitivity problem in the infinite-dimensional case.
\end{remark}
\par \leavevmode \par
\begin{figure}[H]
    \centering
    \includegraphics[width=0.75\textwidth]{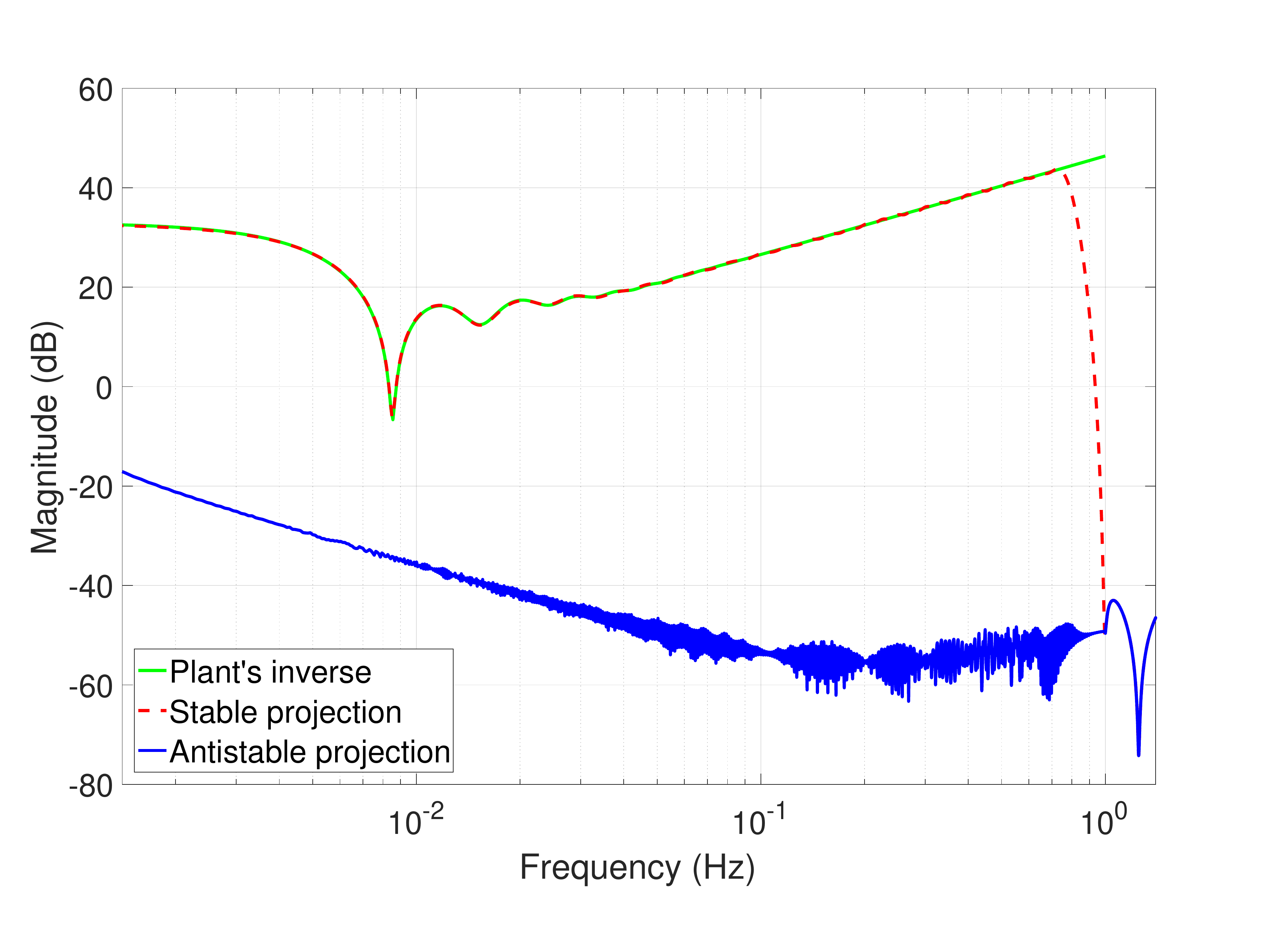}
    \caption{Projection of the inverse of the frequency-response samples from the plant $P$ on the Hardy spaces $\mathcal{H}_2$ and $\overline{\mathcal{H}}_2$: the system is minimum phase.}
    \label{proj_inv_cry}
\end{figure}

 
Step 2 of Algorithm \ref{algo_choixM} consists in doing the same than in Step 1, but for the plant's inverse, in order to determine whether the plant is minimum phase or not. The projection is then performed on the samples $\left\{ \omega_i, P(\imath \omega_i)^{-1}\right\}$. The result is visible on Figure \ref{proj_inv_cry}: the stable projection of the plant's inverse fits the inverse of the plant's frequency-response samples. Consequently, the plant is minimum phase.

\par \leavevmode \par
\begin{figure}[H]
    \centering
    \includegraphics[width=0.75\textwidth]{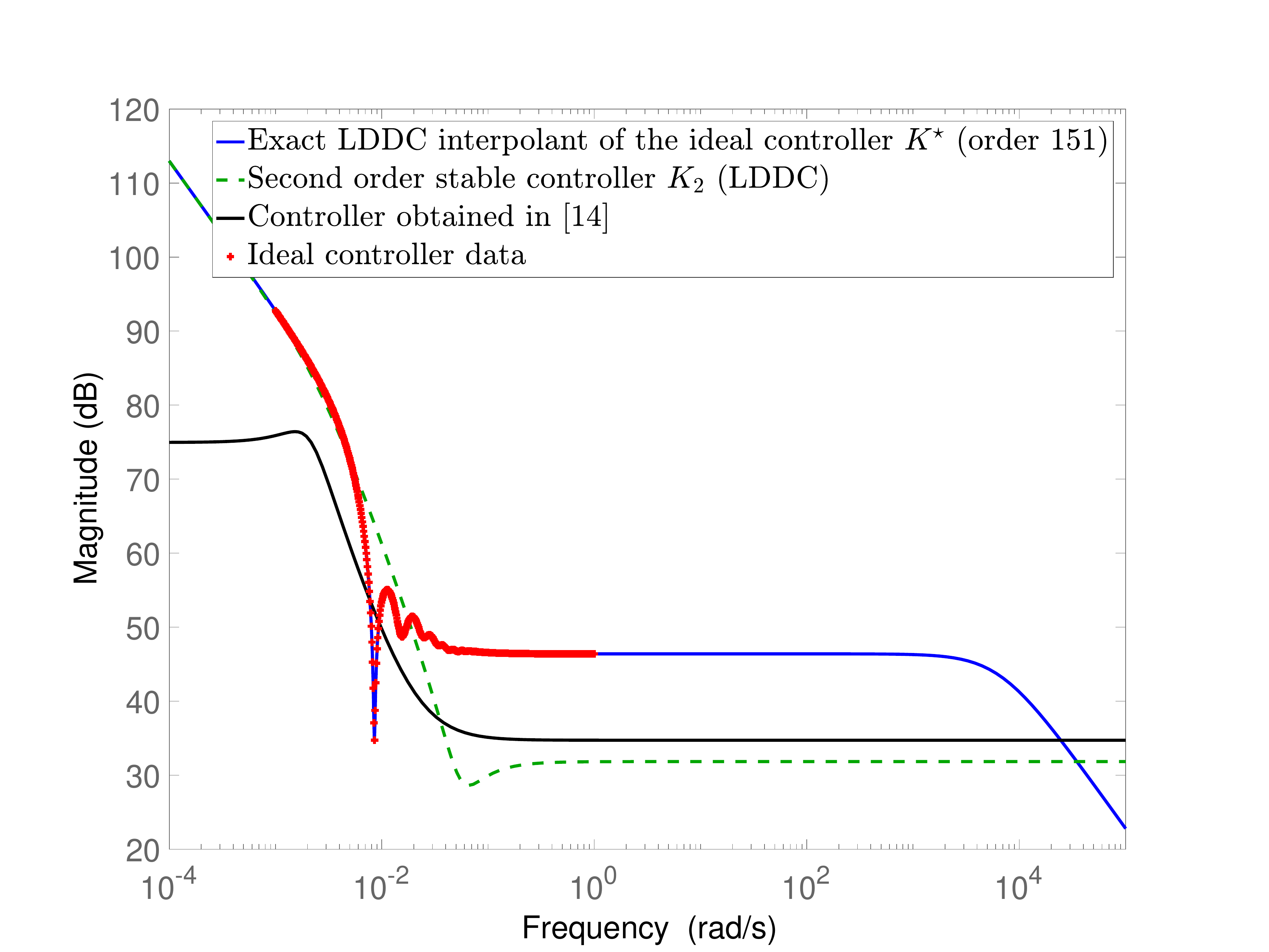}
    \caption{Identification of the controller using the LDDC framework: the minimal realization of the ideal controller has 151 states and is stable. It is reduced to a second-order controller $K$, see \eqref{K_cry}.}
    \label{idK_cry}
\end{figure}

Finally, an achievable reference model is selected according to Step 3 of Algorithm \ref{algo_choixM}. The initial stable reference model $M$ is a first order transfer function:
$$M(s)=\frac{1}{1+\tau s}, \ \tau=1s.$$
Since the plant is unstable and minimum-phase, the achievable reference model is chosen as $M_f=1-(1-M)B_p$, with $B_p$ defined according to the estimated RHP poles of the plant, see Table \ref{RHPpoles}. The LDDC algorithm \cite{kergus2018} is then applied. The identification of the controller is visible on Figure \ref{idK_cry}. The controller is reduced to a second-order model in order order to be compared with the one obtained in \cite{apkarian2017structured}, its expression is given in \eqref{K_cry}. 

\begin{equation}
    K_2(s)=\frac{39.084 (s^2 + 0.04163s + 0.003132)}{s (s+0.002751)}
    \label{K_cry}
\end{equation}

\par \leavevmode \par
\begin{figure}[h]
    \centering
    \includegraphics[width=0.75\textwidth]{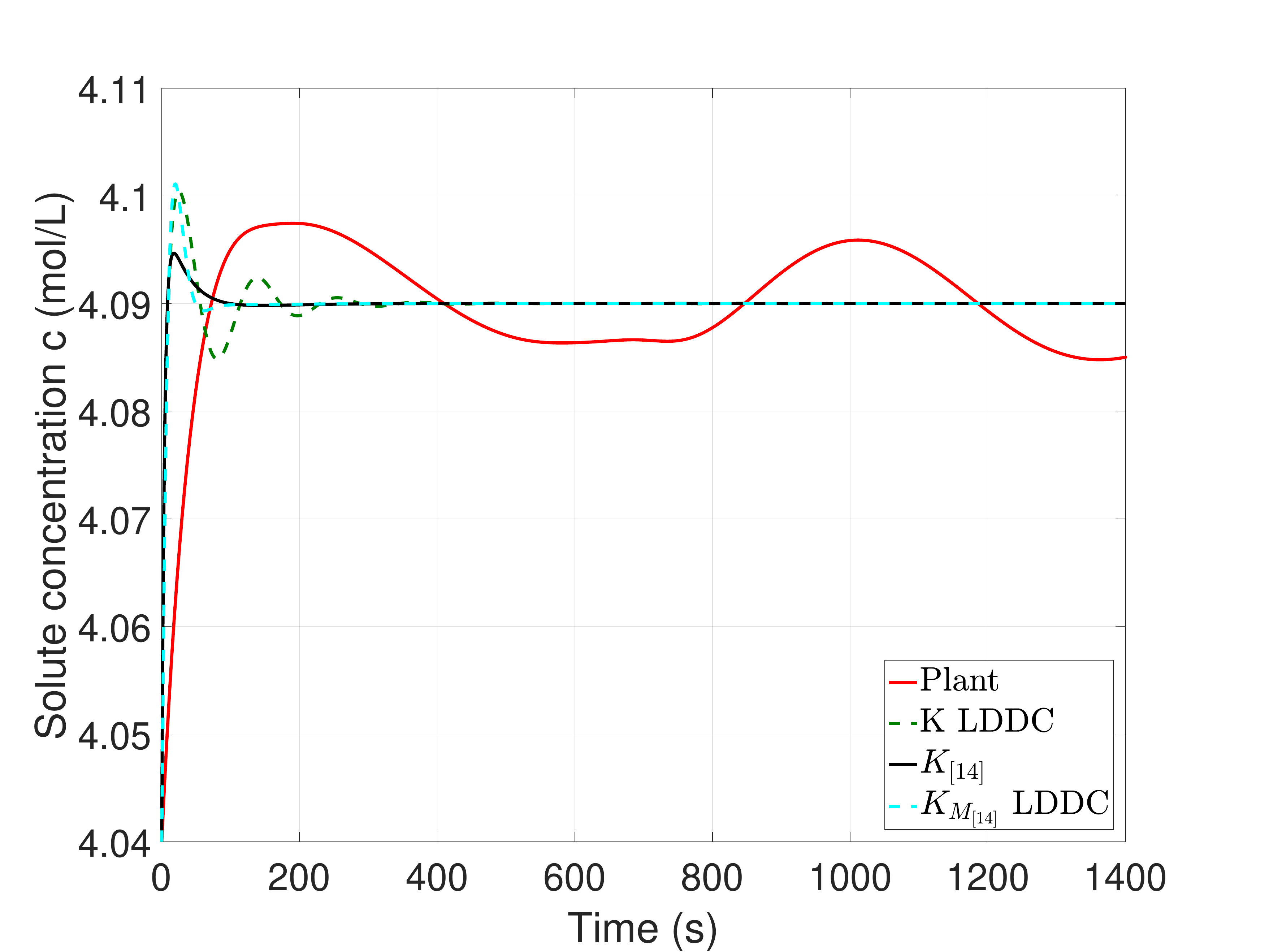}
    \caption{Simulation of the passage to a new steady state for the closed-loops built with the controllers obtained by the LDDC algorithm \cite{kergus2018} using two different reference models and in \cite{apkarian2017structured}.}
    \label{simul_cry}
\end{figure}

The finite-difference model is then used to simulate the closed-loop behaviour in time-domain. The identified controller is compared to the one obtained in \cite{apkarian2017structured}, denoted $K_{[14]}$:
\begin{equation}
    K_{[14]}(s)=\frac{54.47s^2+2.317s+0.02446}{s^2+0.002033s+4.374e-6}.
    \label{K14}
\end{equation}
The results are visible on Figure \ref{simul_cry}. The controller obtained in \cite{apkarian2017structured} allows to reach the desired steady-state faster with a less important overshoot. These better might be explained by two reasons: (i) model-reference control is really limitating when it comes to the expression of the closed-loop specifications and (ii) the reduction of the ideal controller degrades the closed-loop control performances. 

\par \leavevmode \par
\begin{figure}[H]
    \centering
    \includegraphics[width=0.75\textwidth]{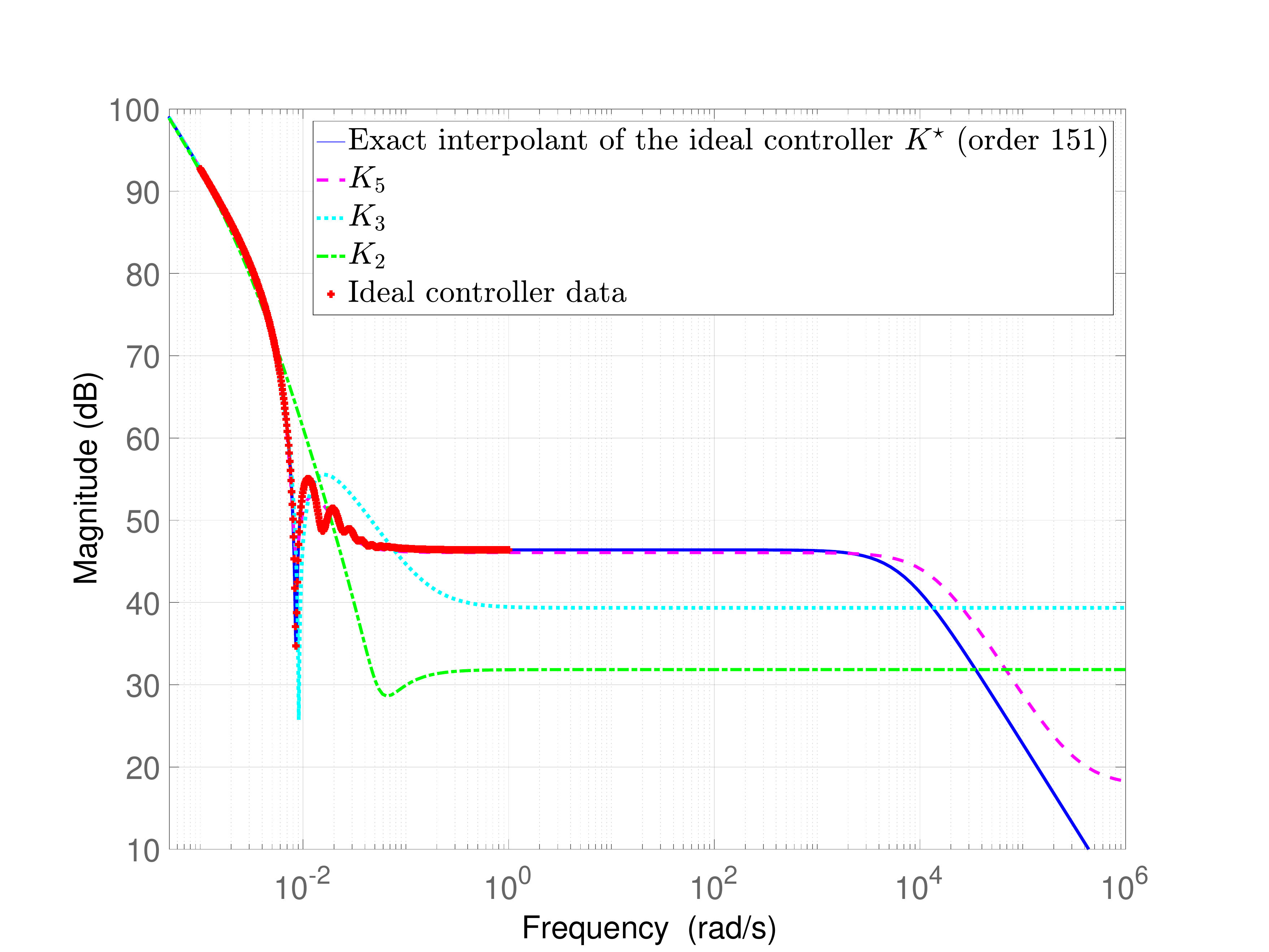}
    \caption{Reduction of the ideal controller to different orders (2,3 and 5). The corresponding controllers are denoted $K_2$, $K_3$ and $K_5$ respectively.}
    \label{reduction_Lddc}
\end{figure}

\begin{figure}[h]
    \centering
    \includegraphics[width=0.75\textwidth]{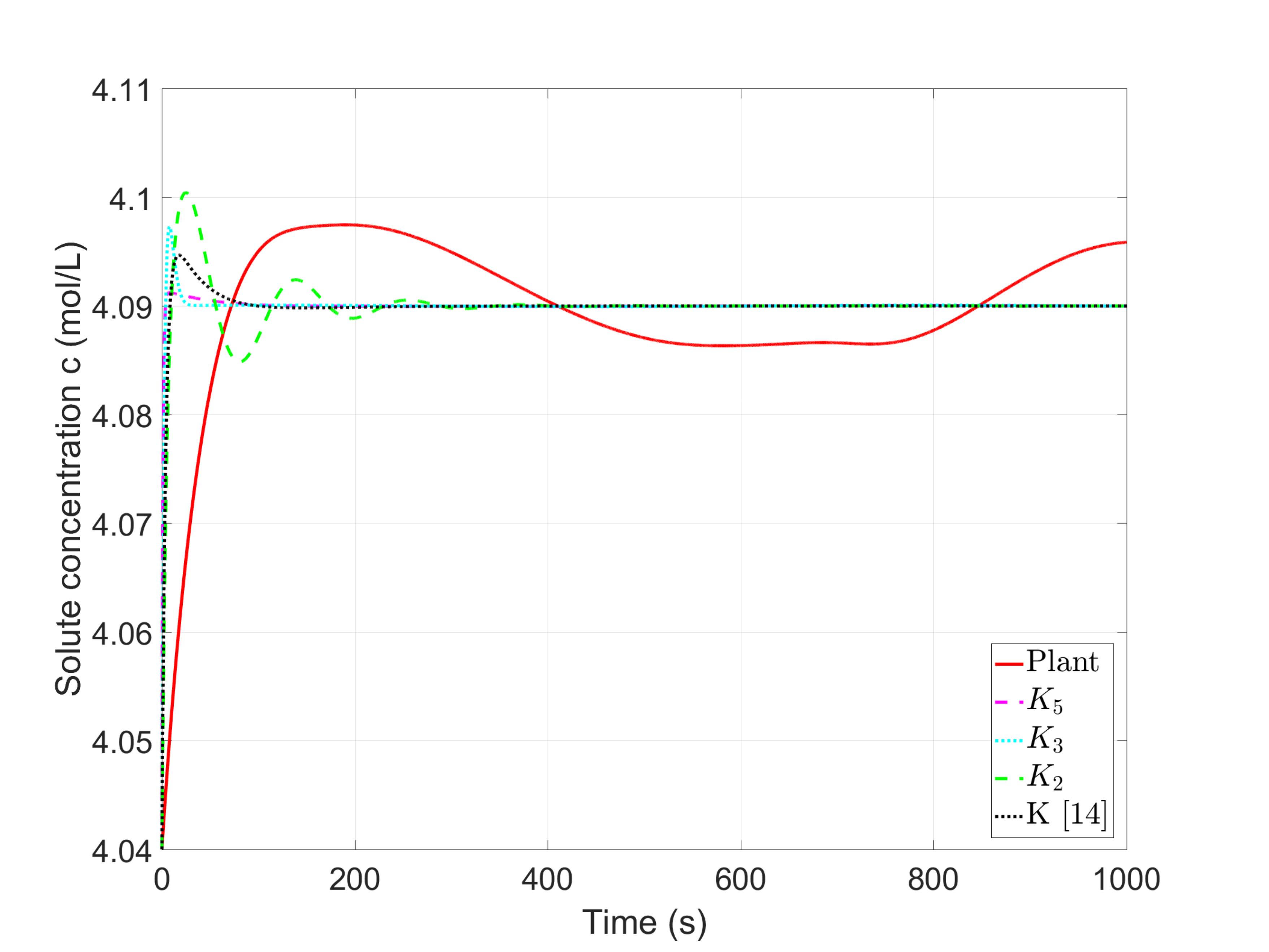}
    \caption{Simulation of the passage to a new steady state for the closed-loops built with the controllers $K_2$, $K_3$ and $K_5$, obtained by the LDDC algorithm \cite{kergus2018} reducing their order to 2, 3 and 5 respectively. They are compared to the controller $K_{[14]}$ obtained in \cite{apkarian2017structured}, see \eqref{K14}. }
    \label{simu_temp_order_lddc}
\end{figure}

 
First, let us investigate (i) the influence of the reduction of the ideal controller on the closed-loop control performances. As shown on Figure \ref{reduction_Lddc}, the more the order of the identified controller is important, the more it will fit the frequency-response of the ideal controller. This is also visible on the time-domain simulations given on Figure \ref{simu_temp_order_lddc}, when simulating the passage to a new steady-state. A high order controller is more likely to give the desired closed-loop behaviour, specified by the reference model $M_f$.

\par \leavevmode \par 
Finally, let us investigate (ii) the influence of the specifications. In \cite{apkarian2017structured}, the specifications are given as frequency weightings functions, giving more freedom to the desired closed-loop behaviour. To underline this aspect, the closed-loop $M_{14}$ reached by the controller $K_{[14]}$ obtained is \cite{apkarian2017structured} is taken as reference model. $M_{14}$ has been computed using the finite-difference model $P_{502}$. It leads to the identification of the second-order controller $K_{M_{14}}$:
\begin{equation}
    K_{M_{14}}(s)=\frac{28.282 (s+0.08652) (s+0.00982)}{(s^2 + 0.001967s + 4.192e-06)}.
    \label{KM14}
\end{equation}
$K_{M_{14}}$ gives a better response time and a more important overshoot than $K_{[14]}$. However, the closed-loop performances induced by $K_{M_{14}}$ are much closer to the ones obtained by $K_{[14]}$ than the ones obtained using $M_f$ as a reference model.

\par \leavevmode \par 
To sum up, the first important aspect in order to obtain good closed-loop performances is the reduction of the ideal controller. The second one is that, for a given controller order, the choice of the reference model remains a critical step: choosing one desired transfer, even achievable, is limiting compared to robust specifications using frequency weightings.

\par \leavevmode \par 
To compare with a model-based method, in \cite{vollmer2001h}, an infinite-dimensional model-based $\mathcal{H}_\infty$ controller synthesis is performed on this very same application. The obtained controller is irrational and a reduction step is needed, which can be quite complicated. On this use-case, a reduced 8th order controller is obtained in \cite{vollmer2001h}. On the other side, applying a structured model-based technique such as hinfstruct relies on an approximation of the irrational system, here the finite-difference model $P_{502}$, and is therefore time-consuming due to the complexity of the model.
\par \leavevmode \par 
For these reasons, data-driven control techniques are particularly indicated in this case. In \cite{apkarian2017structured}, the considerations regarding the limitations due to the RHP poles and zeros of the plant are known through an initial stabilizing controller. Stability is guaranteed by the algorithm thanks to a test on the winding number. However, this method requires to build a fine frequency grid on which samples of the frequency-response of the plant are assumed to be available. Furthermore, the control design relies on iterative non-smooth optimization, which can be time-consuming and is sensitive to the considered initial stabilizing controller.
\par \leavevmode \par 
On the other side, the main strength of the LDDC algorithm is its simplicity. It is a one shot technique, it does not make strong assumptions and does not depend on an initial stabilizing controller.

\section{Application to an hydroelectricity generation channel}
\label{appli_EDF}
The second application is an industrial problem provided by the French power producer EDF (Electricit\'e de France). EDF uses water resources to generate green energy with run-of-the-river power plants. They rely on open-channel hydraulic systems that are non-linear and which dynamic depends on the operating point. Here, for simplicity we will consider one single operating point only.
\par  \leavevmode \par 
Their physical model requires partial differential equations (namely Saint-Venant equations). In \cite{EDF}, a new irrational transfer function is proposed for open channels to represent the level-to-flow variations for any operating point. It is the solution of Saint-Venant equations under many assumptions. The system has two inputs, the entering and the outgoing flows $q_e$ and $q_s$, and one output, the water depth $h$. The transfer is given by: 
\begin{equation}
\begin{array}{ccl}
h(x,s,Q_0) & =  & G_e(x,s,Q_0)q_e(s)+G_s(x,s,Q_0)q_s(s) \\
 & = & P(x,s,Q_0)\left[ \begin{array}{c} q_e \\ q_s \end{array}\right]
 \end{array},
 \label{EDF_eq}
\end{equation}
where
$$\begin{array}{r@{=}l}
G_e(x,s,Q_0) &   \frac{\lambda_1(s)e^{\lambda_2(s)L+\lambda_1(s)x}-\lambda_2(s)e^{\lambda_1(s)L+\lambda_2(s)x}}{B_0s(e^{\lambda_1(s)L}-e^{\lambda_2(s)L})}\\
G_s(x,s,Q_0) &  \frac{\lambda_1(s)e^{\lambda_1(s)x}-\lambda_2(s)e^{\lambda_2(s)x}}{B_0s(e^{\lambda_1(s)L}-e^{\lambda_2(s)L})}
\end{array}$$
where $x$ is the position of the measurement point on the channel, $Q_0$ the nominal flow, $L$ the length of the open channel. $B_0$, $\lambda_1(s)$ and $\lambda_2(s)$ depend on the canal configuration and the nominal flow (see \cite{EDF}). 
\par \leavevmode \par
\begin{figure}
    \centering
    \includegraphics[scale=0.4]{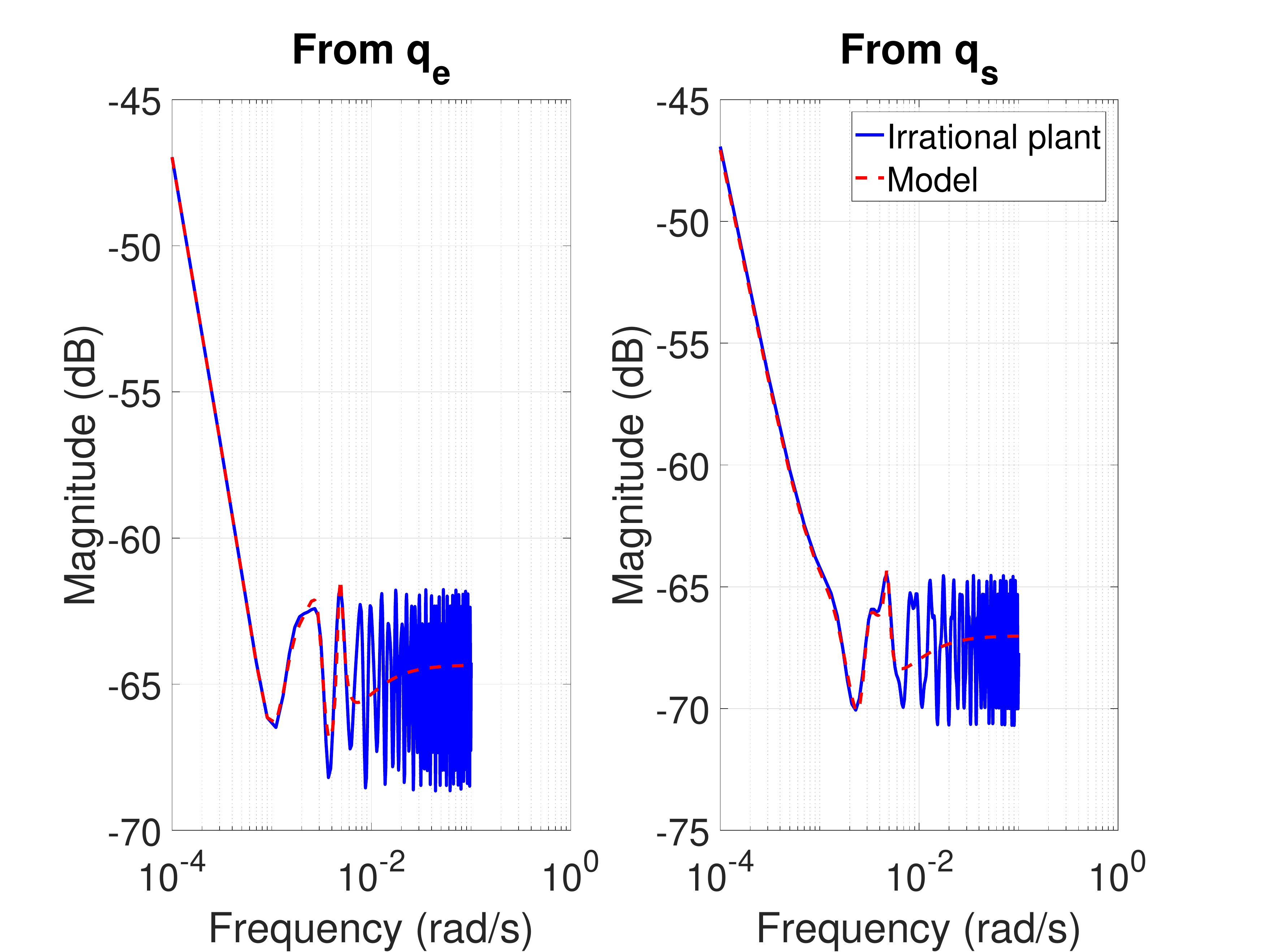}
    \caption{System dynamic for nominal flow $Q_0=1400\ m^3.s^{-1}$: original plant in solid blue and approximation obtained in \cite{EDF} in dashed red.}
    \label{transfertEDF_parametrique}
\end{figure}

 
The system, which dynamic is visible in Figure \ref{transfertEDF_parametrique}, is extremely slow, has a delay behavior and a pole in limit of stability. Moreover, it has an infinite number of poles since the transfer function is irrational. 
\par \leavevmode \par 
The system have been approximated in \cite{EDF} by a 8th order rational transfer function depending on the nominal flow with input time delays: $\tau_e\simeq500s$ and $\tau_s\simeq1500s$, on $q_e$ and $q_s$ respectively. This approximation is shown in Figure \ref{transfertEDF_parametrique} (dashed red). This model will be used to evaluate the performances of the identified controller.
\par \leavevmode \par 
The input flow $q_e$ in the open channel is seen as a disturbance (rain for example). The objective is to maintain the water depth to avoid flooding in the area. To this aim, the command signal is the output flow $q_s$. Therefore, only the transfer $G_s$ between the output flow $q_s$ and the water depth $h$ is considered, see \eqref{EDF_eq}.
\par \leavevmode \par 
In this example, the frequency approach is interesting since the system is represented by an irrational transfer function. Therefore, one cannot have a time-domain simulation. However we still can estimate samples of the frequency response of the system $\{\omega_{i},\Phi_{i}\}, \ i=1 \ldots N$, from which the ideal controller's frequency response can be deduced. The samples of the frequency response $\Phi_{i}=G_s(\imath\omega_{i}), \ i=1 \ldots N$ are extracted from the irrational transfer function $G_s$, for $N=500$ linearly spaced frequencies between $10^{-4}$ and $10^{-1}$ rad.s$^{-1}$.
\par\leavevmode\par
\begin{figure}[h]
    \centering
    \includegraphics[width=0.75\textwidth]{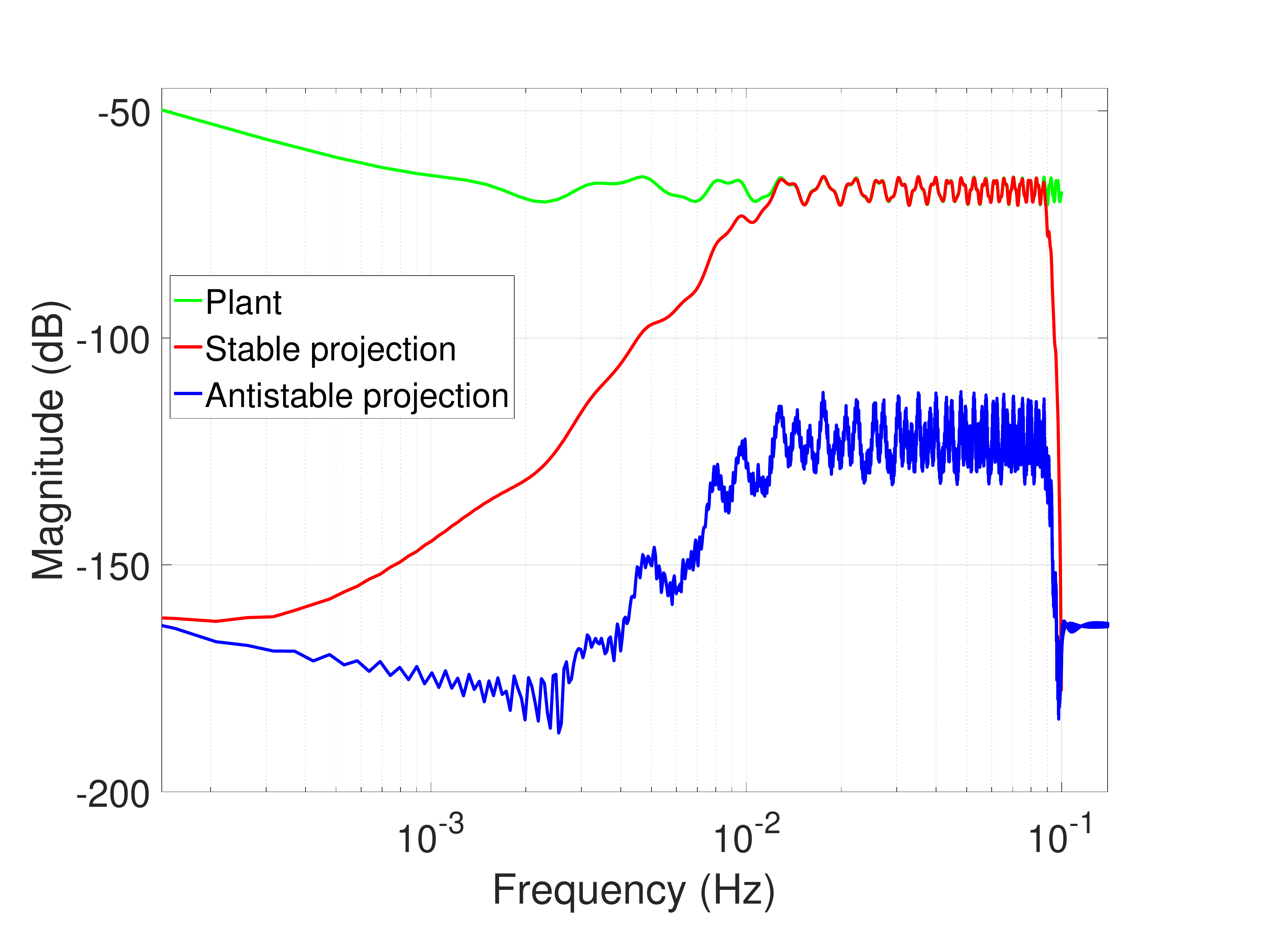}
    \caption{Projection of the plant's data: the only instability is the integrator.}
    \label{proj_EDF}
\end{figure}
\begin{figure}[h]
    \centering
    \includegraphics[width=0.75\textwidth]{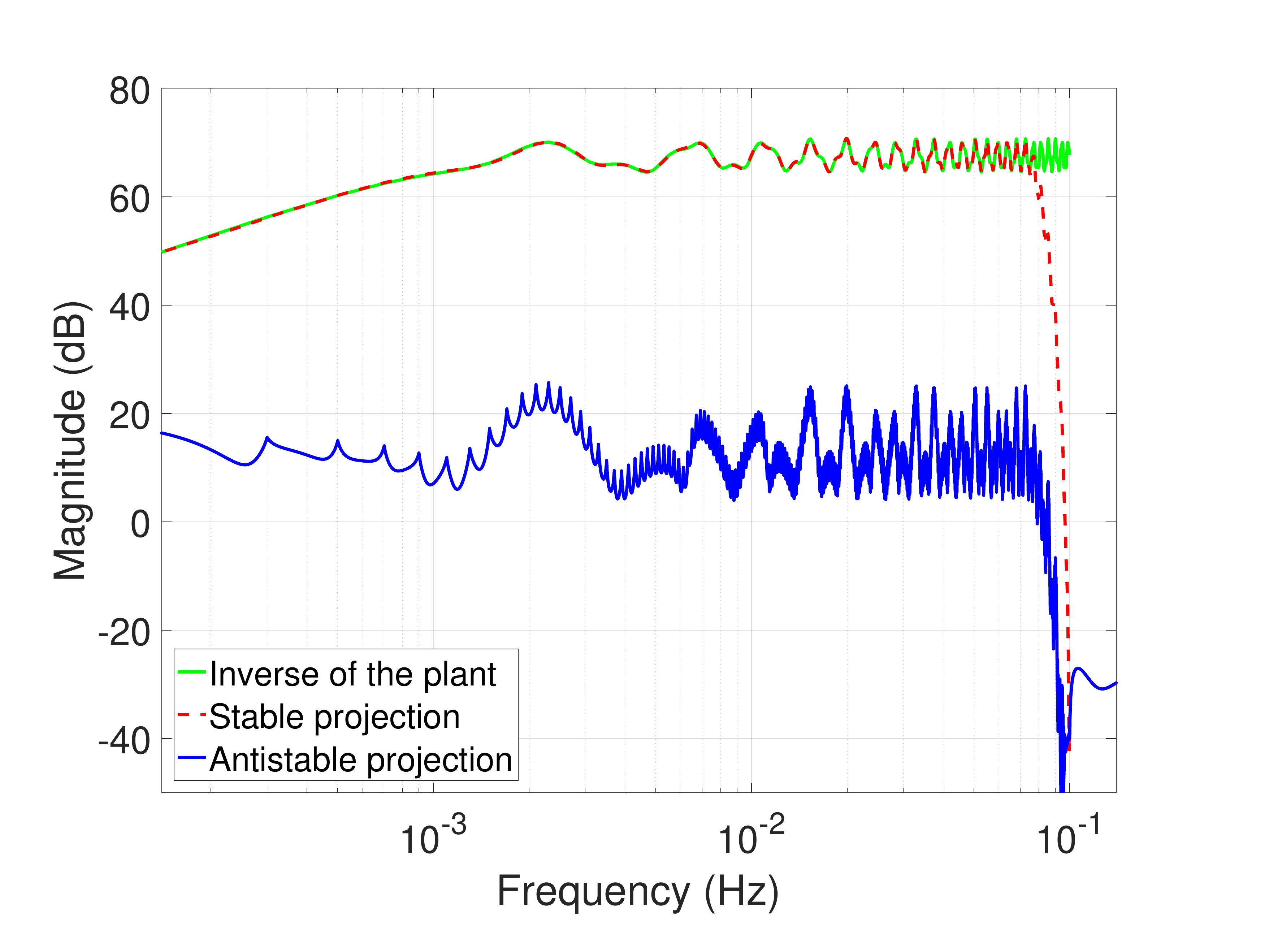}
    \caption{Projection of the inverse of the plant's data: the system is minimum phase.}
    \label{inv_proj_EDF}
\end{figure}

As for the previous application, the first step of the proposed method consist in determining the nature of the system thanks to the projection of its frequency-response measurements. The results of steps 1 and 2 of Algorithm \ref{algo_choixM} are visible on Figures \ref{proj_EDF} and \ref{inv_proj_EDF} respectively. 
\par \leavevmode \par 
Due to the presence of an integrator, the data of the plant is filtered by a bandpass filter for the stability analysis, see \cite{cooman2018model} for further explanations. Figure \ref{proj_EDF} shows that, except the integrator, the plant has no unstable poles. The mismatch between the projection and the plant's data is due to the use of the bandpass filter. According to Figure \ref{inv_proj_EDF}, the plant is minimum phase: the stable projection fits the inverse of the plant's data.

\par \leavevmode \par 
Therefore, the only constraint that the reference model should satisfy is $M(0)=1$, which would have been respected anyway to have zero tracking error. The objective is to stabilize the system and to obtain a faster dynamic. The reference closed-loop $M$ is chosen to be a second order continuous transfer function:
\begin{equation}
    M(s)=\frac{1}{1+\frac{2\xi}{\omega_0}s+\frac{s^2}{\omega_0^2}},
    \label{M_EDF}
\end{equation}
with $\omega_0=10^{-4} rad.s^{-1}$ and $\xi=1$. It satisfies $M(0)=1$.

\par \leavevmode \par 
The frequency response of the ideal controller $K(\imath\omega_{i})$, which exactly provides the desired closed-loop behavior dictated by $M$ when placed in the closed-loop, is obtained as follows:
$$\forall i=1\ldots N, \ K(\imath\omega_{i})=\frac{M(\imath\omega_{i})}{\Phi_{i}-M(\imath\omega_{i})\Phi_{i}} .$$

\par \leavevmode \par
\begin{figure}[H]
    \centering
    \includegraphics[width=0.75\textwidth]{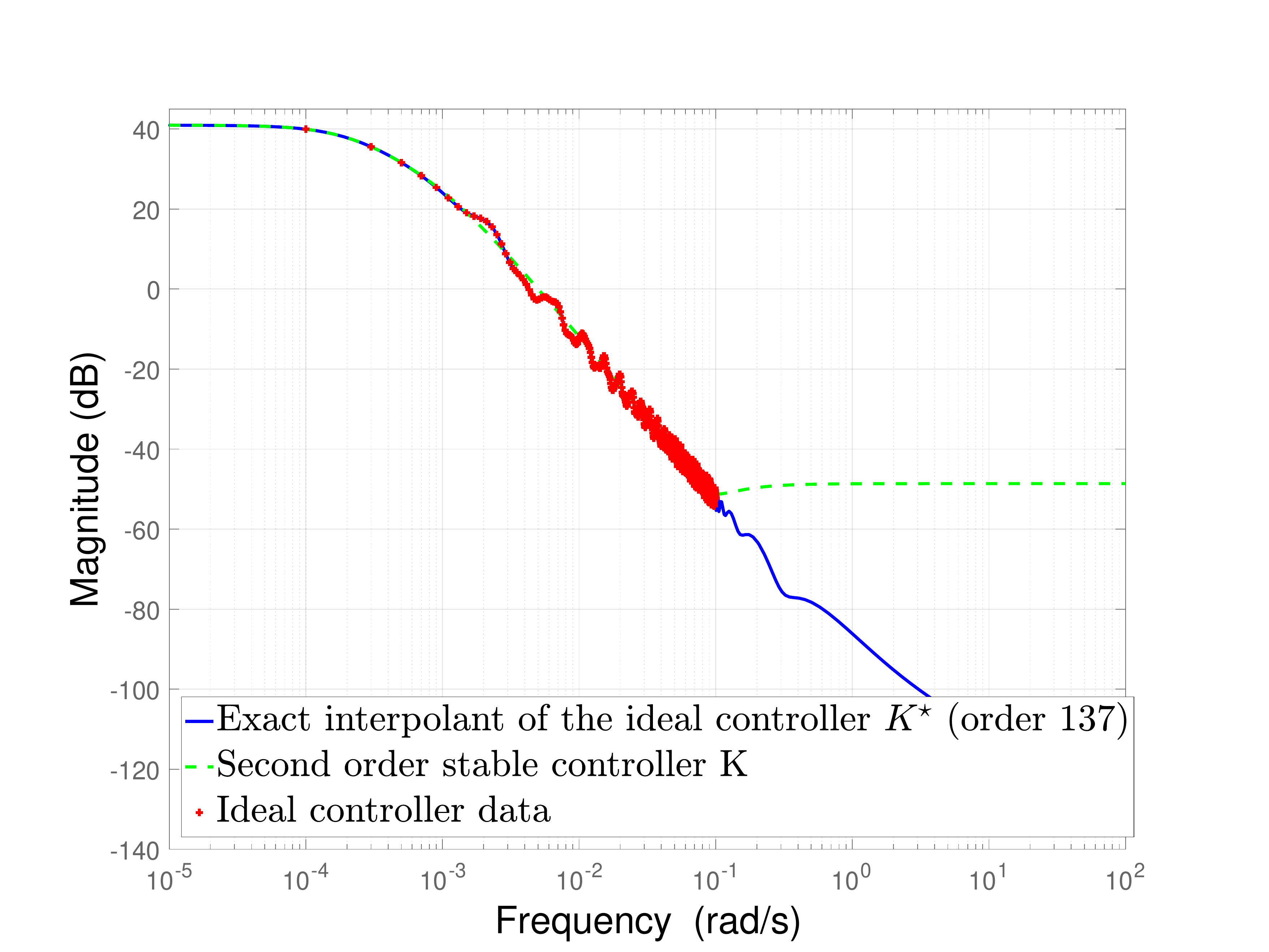}\\
    \caption{Identification of the ideal controller (red dots): minimal realization of order 137 (solid blue) and reduced 2nd order LDDC controller(dashed blue).}
    \label{id_K_EDF}
\end{figure}

The result of the identification step is given in Figure \ref{id_K_EDF}. The minimal realisation of the ideal controller is of order 137 and is stable. It is then reduced to a stable second order controller:
$$K(s)=\frac{3.3492e-05 (s^2 - 0.0663s + 0.007729)}{(s+2.001e-05) (s+0.001161)}.$$

\par \leavevmode \par
\begin{figure}[H]
    \centering
    \includegraphics[width=0.75\textwidth]{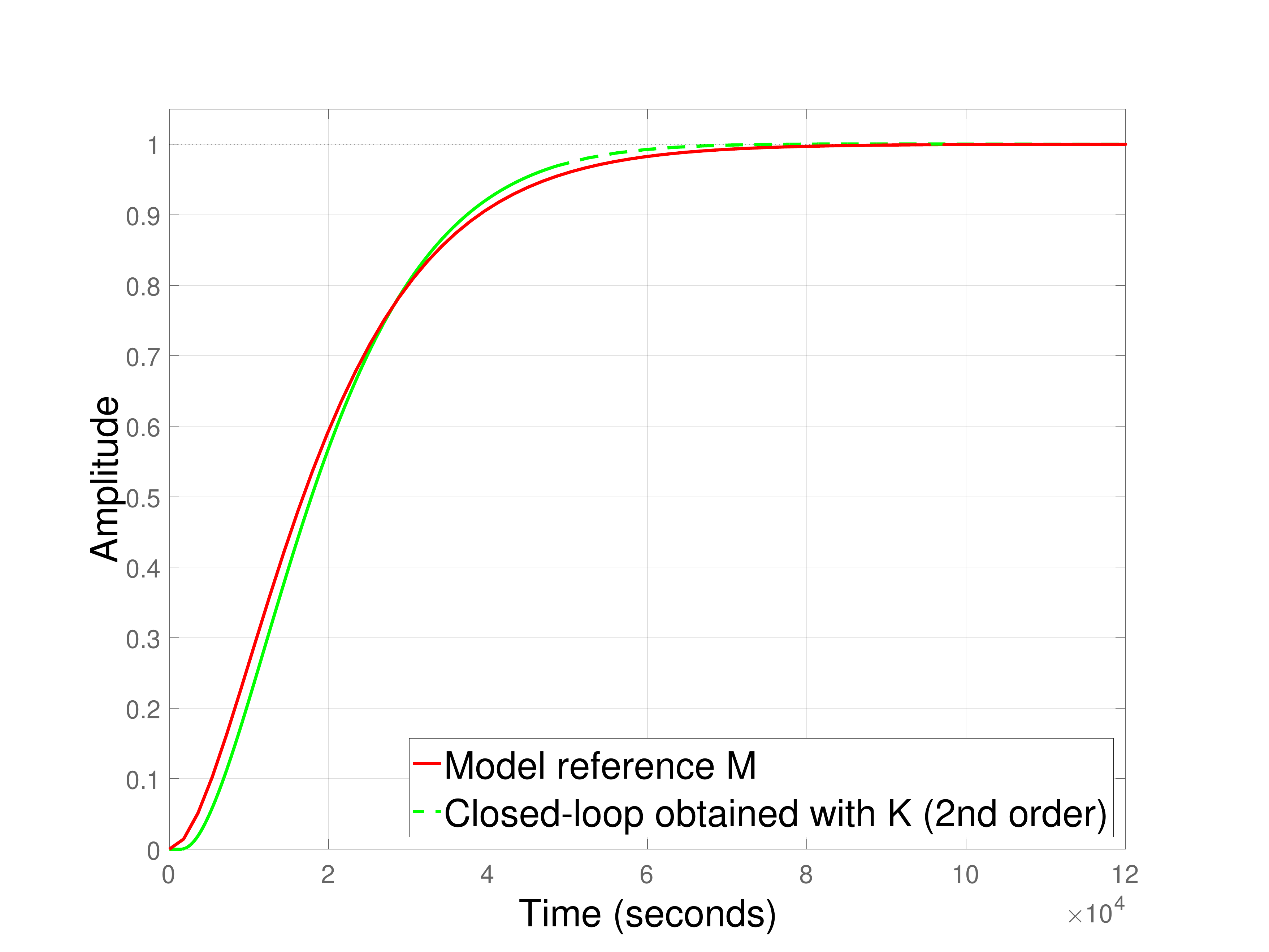}
    \caption{Step response of the closed loop using the 2nd order LDDC controller (dashed green) and of the objective transfer (solid red).}
    \label{simul_temp_EDF_1}
\end{figure}

\begin{figure}[H]
    \centering
    \includegraphics[width=0.75\textwidth]{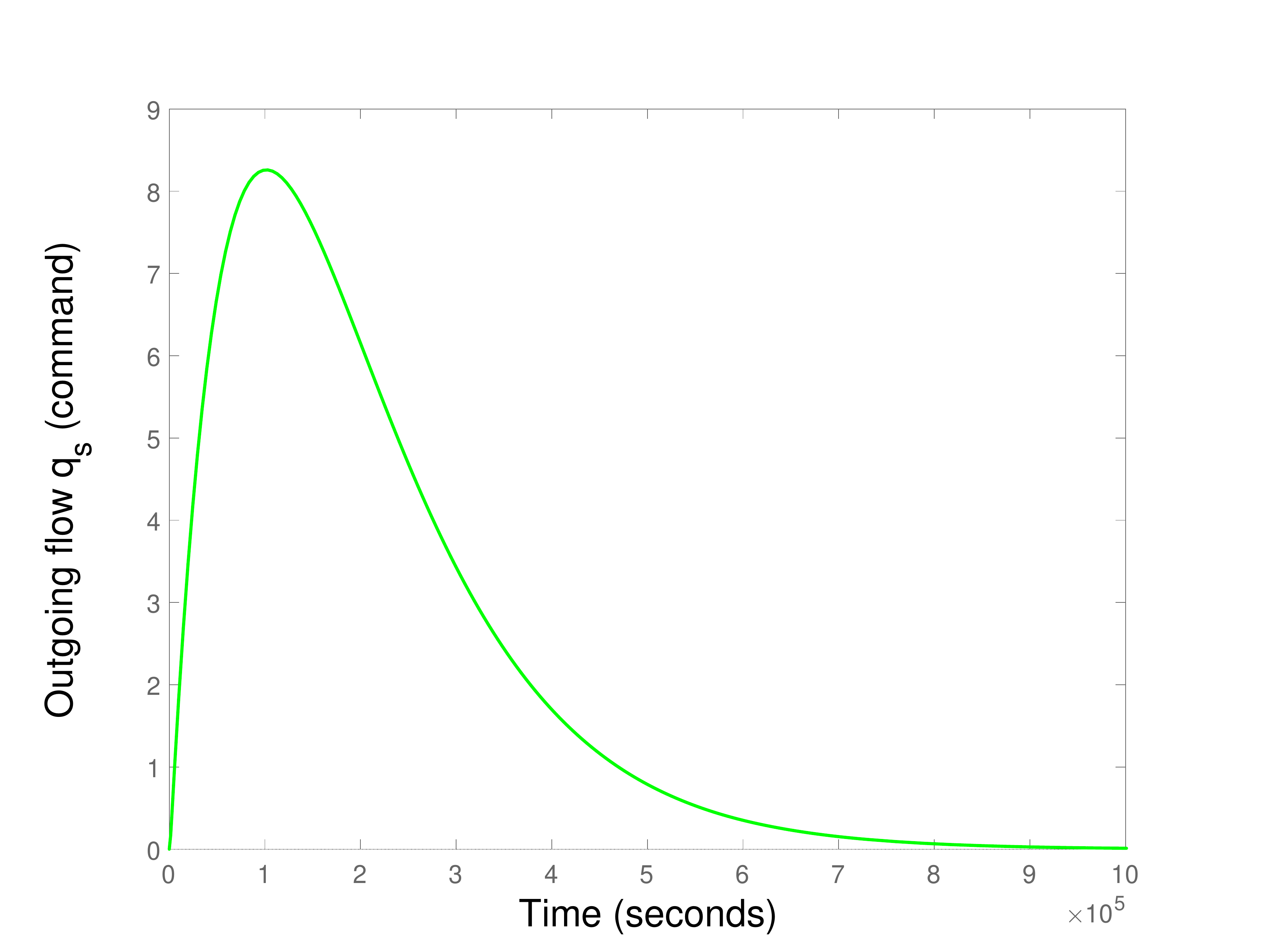}
    \caption{Evolution of the outgoing flow $q_s$ when applying a step on the closed-loop obtained with the identified 2nd order controller $K$.}
    \label{simul_temp_EDF_u}
\end{figure}

 
Since we had no access to the EDF simulator, the 8th order rational transfer function of \cite{EDF} is used to simulate the closed-loop with the 2nd order controller obtained by the LDDC. The results are shown in Figure \ref{simul_temp_EDF_1}: the resulting closed-loop achieve a response time of $4.84\times10^5$s (134.4 hours) with no overshoot, while the system naturally has a response time of $2.47\times10^{13}$s. The closed-loop dynamic is almost the objective one. The command signal is shown on Figure \ref{simul_temp_EDF_u}: it is reasonable, the maximum flow variation is around $8.3m^3.s^{-1}$, which is in the acceptable range for this application with a controller of order 2.

\par \leavevmode \par
\begin{figure}[H]
    \centering
    \includegraphics[width=0.75\textwidth]{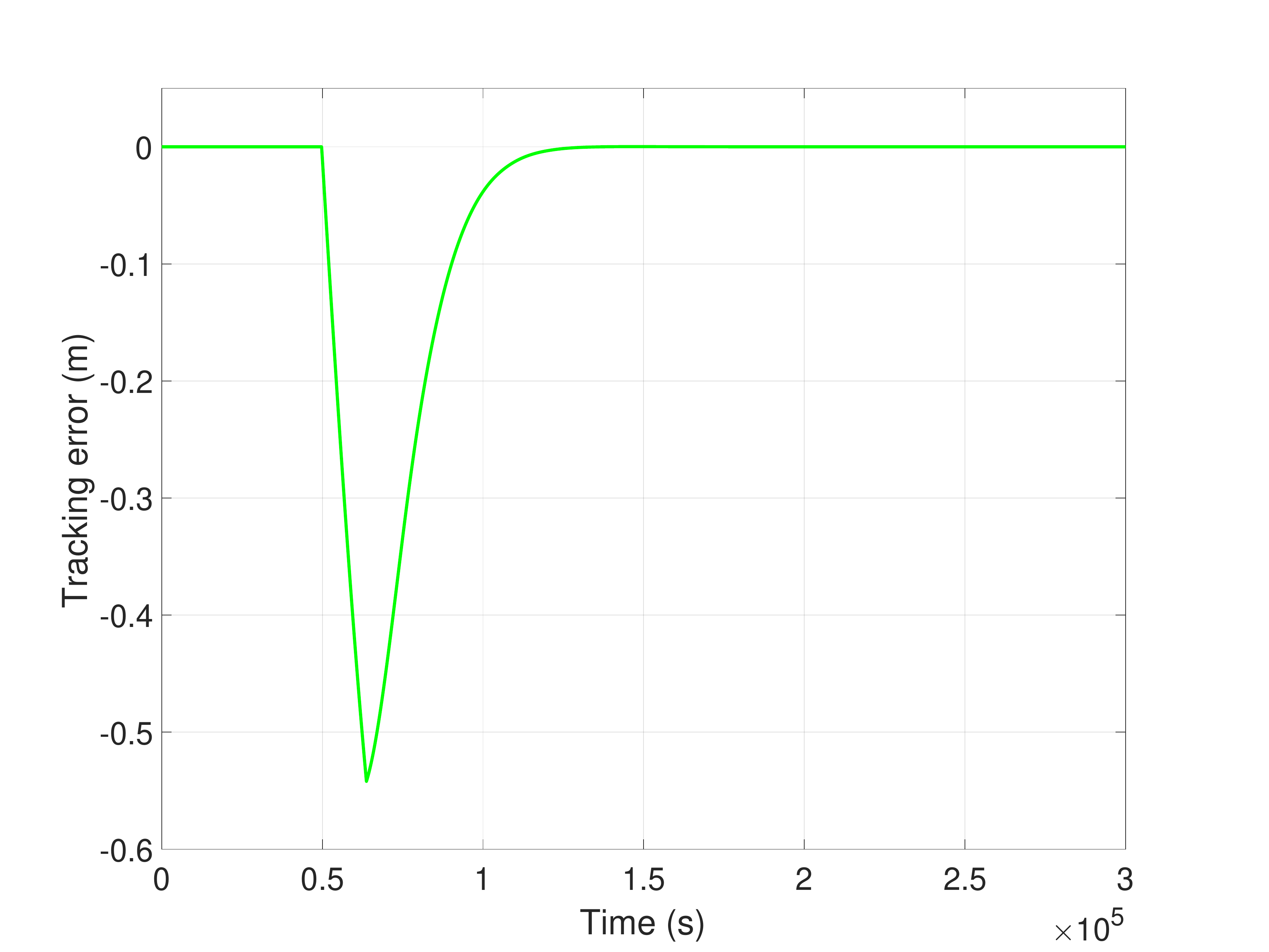}
    \caption{Performance of the controlled closed-loop in terms of disturbance rejection.}
    \label{EDF_dist}
\end{figure}

 
Finally, the performances in terms of disturbance rejection are shown on Figure \ref{EDF_dist}: an input flow of $100m^3/s$ during four hours is considered. This disturbance would increase the water depth of $0.65m$ if not rejected. Figure \ref{EDF_dist} represents the tracking error when this disturbance is applied on the stabilized closed-loop: the water depth increases of $0.54m$ instead of the $0.65m$ without the controller. Finally, the disturbance is completely rejected $0.8\times 10^{5}s$ (2.3 hours) after its application. During the remaining time of the disturbance application, the disturbance does not affect the closed-loop system.

\par \leavevmode \par
It should be noted that the closed-loop could be faster by increasing the frequency $\omega_0$ of the reference model $M$, see \eqref{M_EDF}. However, taking $\omega_0=10^{-4}rad.s^{-1}$ (instead of $10^{-5}rad.s^{-1}$), leads to the identification of a second-order controller giving an oscillatory behaviour in closed-loop. The consequence is a significant overshoot, which is not acceptable in this application. Furthermore, the command signal $q_s$ rises to $80m^3/s$ when applying a step on the corresponding closed-loop.

\par \leavevmode \par 
In this application, the LDDC method is appealing since it does not require to simulate the complex system described by an irrational transfer function to obtain time-domain data. Only samples of the frequency response of the plant are needed, which can be estimated directly from the irrational transfer function. Moreover, one should notice that controlling such an infinite order model is also quite challenging even for model-based methods. A interesting perspective would be to try this controller on the EDF simulator instead of using the approximate model to validate the performances.

\section{Conclusions}
\label{conclusion}
In this appendix, two additional examples illustrate the method proposed in \cite{kergus2019} to select an achievable reference model for data-driven control purposes. The first one is a continuous crystallizer and the second one is an open-channel for hydroelectricty generation. These applications are representative of the class of systems for which data-driven control techniques is more appealing than model-based ones. Indeed, their model, which are irrational, are too complex for model-based control. 
\par In both cases, the data-driven stability analysis proposed in \cite{cooman2018model} allows to draw the right conclusion concerning the nature of the plant. However, the second example considered in Section \ref{appli_EDF} shows that integrators are not handled by this technique. In Section \ref{appli_cry}, the continuous crystallizer is shown to be unstable and its unstable poles correspond to the ones previously found in the literature.
\par The techniques presented in \cite{kergus2019} and recalled in Section \ref{algo} is then used to select an achievable reference model. The resulting ideal controllers are stable and, in the case of the continuous crystallizer, do not compensate the plant's instabilities. The LDDC algorithm is then used to identify reduced-order controllers.
\par As shown by the application of the LDDC algorithm on the continuous crystallizer, the proposed method does not compensate the fact that choosing a reference model constitutes a very limited specifications requirement compared to robust specifications. Further work will treat the choice of a reference model in the multivariable case: the problem becomes a tangential interpolation one. Another outlook would be to investigate the possibility to tune the reference model to have a better control on the corresponding closed-loop performances.



\vspace{-0.3cm}
\bibliographystyle{unsrt}
\bibliography{biblio}


\end{document}